\documentclass[12pt]{article}
\usepackage{enumerate}
\usepackage{amsmath}
\usepackage{hyperref} 
\hypersetup{colorlinks, breaklinks, urlcolor=blue, linkcolor=blue} 
\usepackage{slashed}
\usepackage[utf8]{inputenc}
\usepackage{authblk}

\begin{document}

\title{Remarks on Top-philic $Z^\prime$ Boson Interactions with Nucleons}
\author{Frederick S. Sage, Jason N.E. Ho, T.G. Steele, and Rainer Dick}
\affil{
Department of Physics and
Engineering Physics, University of Saskatchewan, Saskatoon, SK,
S7N 5E2, Canada
}

\date{}

\maketitle

\abstract{This article provides the calculation of an effective vertex function between a nucleon and a $Z^\prime$ boson that couples preferentially to either the top quark or the third generation of fermions, for the purpose of calculating vector-portal dark matter nuclear recoil cross sections. Mixing effects between the new gauge group $U(1)^\prime$ and the Standard Model hypercharge group $U(1)_Y$ are taken into account. Contributions to the $U(1)^\prime$ nucleon current from heavy quarks are quantified using the heavy quark expansion. Also taken into account are contributions from the 1-loop $Z^\prime$-gluon interactions and mixing-induced contributions from the light quarks in the nucleon. We find that, for reasonable values of the $U(1)^\prime$ gauge parameter, contributions from the light quarks dominate despite being mixing-suppressed. It is shown that this holds for most models even if mixing effects do not appear at tree level. Contributions from the heavy quarks and gluons are suppressed by $1/m_Q^2$ and possibly also by momentum in the low momentum transfer limits relevant for dark matter direct detection. We discuss under which conditions the subdominant terms become relevant.}

\section{Background and Motivation}

The dark matter problem is one of the most compelling in particle physics, primarily because it provides the strongest evidence we have for the existence of physics beyond the Standard Model (SM). However, dark matter is both nonluminous and nonbaryonic, making it very difficult to detect. This has led to the formation of an extensive experimental program to detect dark matter. The three primary channels for observable interactions between the dark and visible sectors are collider production, indirect detection, and direct detection, with the last of these being the focus of this article.

One of the more well-studied classes of dark matter models are hidden sector models \cite{Pospelov:2007mp}. These models include interactions between the dark and visible sectors only through certain mediator fields, the most frequently considered of which are the mass dimension four `portals.' Of these, the vector portal involves interactions mediated by the exchange of a massive vector boson, usually associated with some extended gauge symmetry. The most common realization of vector-portal dark matter involves a $U(1)^\prime$ gauge boson as the mediator, usually denoted by $Z^\prime$. Such vector-portal models with a $Z^\prime$ have been well studied \cite{Alves:2013tqa,Jacques:2016dqz,Dudas:2012pb,Chun:2010ve,Cline:2014dwa,Alves:2015mua}. 

A variant of the $Z^\prime$ that has been gaining interest is the $Z^\prime$ that couples preferentially to the top quark (top-philic) \cite{Cox:2015afa,Arina:2016cqj}, or to the third generation of SM fermions (tritophilic) \cite{PhysRevD.91.035025}. Such models have been motivated by both theoretical considerations and by phenomenological concerns. Top-philic models have appeared in supersymmetric frameworks in the past \cite{Beck:2015cga}, as well as in topcolor-assisted technicolor models \cite{Hill:1994hp,Lane:1995gw,Lane:1996ua,Lane:1998qi,Popovic:1998vb}. Tritophilic models naturally appear in some topflavor scenarios \cite{He:1999vp}, and have been explored for some time \cite{Holdom:1994ss,Frampton:1996bs}.

More recently, studies of top-philic vector-portal dark matter have appeared. These models have the advantage of being able to avoid the stringent nuclear recoil exclusion limits \cite{Akerib:2015rjg,Tan:2016zwf} imposed by direct detection experiments while being relatively unconstrained by other detection channels. The direct detection exclusion limits can be avoided if the mediating $Z^\prime$ has highly suppressed interactions with light quarks. For $Z^\prime$ mediated dark matter where the $Z^\prime$ interacts with the up and down quarks, the interaction between the $Z^\prime$ and the nucleon can be expressed as a sum of the $U(1)^\prime$ charges of the appropriate light quarks. For a $Z^\prime$ that couples preferentially to the top or the third generation, the situation is not so straightforward.

In this article, we calculate a coupling between a top-philic or tritophilic $Z^\prime$ boson and the nucleon. This issue has been treated in passing in the literature \cite{Cox:2015afa}, but never in full detail. We proceed along the same lines as the computation of the classic Shifman-Vainshtein-Zakharov Higgs-nucleon coupling \cite{shifman1978,cheng1988}, which involves using the heavy quark expansion \cite{Novikov:1977dq} to take into account the heavy quark content of the nucleon. In Section 2, we describe the general structure of the theories we will consider, and discuss the effects of kinetic mixing. Section 3 is our calculation of the contributions to the effective $U(1)^\prime$ nucleon current. Implications for the direct detection of dark matter are included in Section 4, and Section 5 is a summary of our results and our conclusions.

\section{Top-philic $U(1)^\prime$ Interactions}

Our calculations and results are formulated in terms of a generic $Z^\prime$ model. We assume the $Z^\prime$ is the gauge boson of some local $U(1)^\prime$ symmetry in the gauge eigenstate with a mass $m_{Z^\prime}$. We remain agnostic as to how this mass is generated, making no assumptions about possible UV physics.

In the basic model, only the top quark (or the top and bottom quarks) are charged under the $U(1)^\prime$ gauge group. What follows could also apply to the lepton sector, but we will frame our discussion in terms of quarks to align with our intended application. Despite only certain quarks having $U(1)^\prime$ charge, there will generally be mixing-induced interactions between the $Z^\prime$ and any particle with nonzero SM hypercharge. Kinetic mixing can appear in the tree level Lagrangian through a term $(\epsilon/2) B^{\prime\mu\nu}B_{\mu\nu}$, where $B^{\prime\mu\nu}$ and $B^{\mu\nu}$ are the field strength tensors associated with the $U(1)^\prime$ and SM hypercharge ($U(1)_Y$) gauge fields. Even if no mass mixing or kinetic mixing is explicitly included in the Lagrangian, mixing will be generated by loop order effects as a consequence of some particles being charged under both $U(1)^\prime$ gauge groups \cite{delAguila:1988jz}. Heavy quark loops will generate kinetic mixing (at 1-loop level) on the order of \cite{Cox:2015afa}

\begin{equation}\label{loopmix}
\epsilon \approx \frac{2}{3}\frac{N_c g_{2} g^\prime}{16 \pi^2} \log \left( \frac{\Lambda_{UV}^2}{m_Q^2}\right),
\end{equation}

and mass mixing on the order of

\begin{equation}
\delta m^2 \approx \frac{1}{2}m_Q^2 \frac{N_c g_{2} g^\prime}{16 \pi^2} \log \left( \frac{\Lambda_{UV}^2}{m_Q^2}\right).
\end{equation}

The mass of the heavy quark $Q$ generating the loop is $m_Q$, $g_2$ is the weak hypercharge gauge parameter, and $N_c=3$ is the number of colors. The regulator $\Lambda_{UV}$ is the scale at which new physics enters; which will usually be the symmetry breaking scale at which the $Z^\prime$ mass is generated. Because of these quantum corrections, models can be perfectly top-philic or tritophilic only classically or in approximation.

We frame our mixing and gauge parameters following \cite{Chankowski:2006jk} (see also \cite{Foot:1991kb,Holdom:1985ag}), which leads to a generic $Z^\prime$-fermion interaction current of the form

\begin{equation}\label{mixZ'1}
J^\mu_{Z^\prime} = \left( g^\prime Y^\prime + g_{mix} Y \right)\bar{f} \gamma^\mu f.
\end{equation}

In the above, the fermions are generic. The SM hypercharge is $Y$ and the $U(1)^\prime$ charge is $Y^\prime$. The $U(1)^\prime$ gauge coupling (after the appropriate rotations) is $g^\prime$, and $g_{mix}$ is the mixed gauge coupling as discussed in the reference \cite{Chankowski:2006jk}. The relationship between $g_{mix}$ and the mixing parameters $\epsilon$ and $\delta m$, whether the mixing parameters are loop-induced or exist in the tree level Lagrangian, is dependent on the symmetry breaking pattern by which the $U(1)_Y \times U(1)^\prime$ substructure is reduced to $U(1)_{em}$. We work exclusively with $g_{mix}$ in this article. 

Generally for anomaly-free models the vector and axial charges of the fermions will be written in terms of the $U(1)^\prime$ charges $Y^\prime_{iL,R}$ of the left and right handed components and the $U(1)^\prime$ gauge coupling $g^\prime$ (this holds for any Abelian $U(1)$ gauge group):

\[
J^\mu_{Z^\prime} = \left( g^\prime Y^\prime_L + g_{mix} Y_L \right)\bar{f} \gamma^\mu P_L f + \left( g^\prime Y^\prime_R + g_{mix} Y_R \right)\bar{f} \gamma^\mu P_R f\]
\begin{equation}\label{mixZ'2}
= \bar{f} \gamma^\mu \left( V^\prime_f + A^\prime_f \gamma^5 \right) f.
\end{equation}

As usual, the left and right handed projection operators are $P_L = (1/2)(1-\gamma^5)$ and $P_R = (1/2)(1+\gamma^5)$. We have introduced in the second line of equation \ref{mixZ'2} the mixed vector and axial generalized charges $V^\prime_f$ and $A^\prime_f$. These quantities are compact notation that is convenient for phenomenological calculations, and our results are expressed in terms of these quantities whenever possible.

Three specific types of fermions are of interest to us. First, there are those SM fermions which are charged under $U(1)^\prime$. These are either the top quark in the top-philic scenario, or the top and bottom quarks in the tritophilic scenario. These fields have both nonzero $Y$ and nonzero $Y^\prime$, so both terms in the current are nonzero. Second, there are fermions with SM hypercharge that have zero $U(1)^\prime$ charge that interact with the $Z^\prime$ proportionally to $g_{mix}$. Finally, we consider a generic cold dark matter candidate $\chi$ that has $U(1)^\prime$ charge but no hypercharge. Such fermions have an interaction strength with the $Z^\prime$ governed only by $g^\prime$. Non-fermionic dark matter will have a different form of interaction, but the interaction will still be proportional to $g^\prime$; see \cite{Berlin:2014tja} for more details.

We mention briefly some experimental constraints on $U(1)^\prime$ extensions of this type. Model-independent studies of LHC data \cite{Greiner:2014qna,Kim:2016plm} have reported bounds on top-philic vector resonances that are stringent for a $Z^\prime$ with a mass of $\mathcal{O} (100 \ \mathrm{GeV})$, but that weaken considerably for higher masses. When a gauge field with a non-universal coupling between families is introduced, flavor changing neutral currents (FCNCs) will be introduced as well. This is a strong experimental constraint on the properties of a $Z^\prime$ which couples differently to the first and second generations, but constraints on FCNCs for the third generation are still weak \cite{Gresham:2011dg,Gupta:2010wt}.

\section{Nucleon Couplings}

In this Section, we will compute the various contributions to the $Z^\prime$-nucleon vertex and discuss which provide the dominant contributions. We label light quark fields either by flavor ${ u, d, s}$ or generically as $q$, and interacting heavy quark fields are $Q$. We work with a generic nucleon $N$, though the $Z^\prime$ current generally does not respect isospin invariance. The resulting differences between the proton and neutron couplings will be addressed only when numerical values are required. The nucleon momenta are not equal in general, and we will label them $P^\mu$ and $P^{\prime\mu}$, though they will be suppressed for brevity throughout most of the Section. However, in the end we will be working in the low momentum transfer limit of small $( P^{\prime\mu} - P^\mu )^2 = k^2$, which is equivalent to the forward matrix element limit frequently used in hadronic physics.

We decompose the interaction current $J_{Z^\prime NN}^\mu$ as follows:

\[
J_{Z^\prime NN}^\mu Z^\prime_\mu \bar{\Psi}_N \Psi_N = \sum_{q=u,d, s} Z^\prime_\mu \langle N | \bar{q}\gamma^\mu \left( V^\prime_q + A^\prime_q \gamma^5 \right) q | N \rangle +  \sum_{Q} Z^\prime_\mu \langle N | \bar{Q}\gamma^\mu \left( V^\prime_Q + A^\prime_Q \gamma^5 \right) Q | N \rangle \]
\begin{equation}
+ \langle N | \Gamma_{Z^\prime gg}^{\mu\nu\rho} \left( p_1 ,p_2 \right) Z^\prime_\mu \left( k \right) B_\nu \left( p_1 \right) B_\rho \left( p_2 \right) | N \rangle .
\end{equation}

In order, the terms on the right hand side correspond to the contributions to the nucleon current from the light quarks, the interacting heavy quarks, and the gluons. The light quark term can be written in terms of the neutral current form factors of the nucleon. Making the identification

\begin{equation}
V_q^\prime \langle N | \bar{q}_j \gamma^\mu q_j | N \rangle = V_q^\prime \bar{u} \left[ F_1^j (k^2) \gamma^\mu + i\frac{\sigma^{\mu\nu}q_\nu}{2m_N} F_2^j (k^2) \right] u,
\end{equation}

and

\begin{equation}
A_q^\prime \langle N | \bar{q}_j \gamma^\mu \gamma^5 q_j | N \rangle = A_q^\prime \bar{u} \left[ G_A^j (k^2) \gamma^\mu \gamma^5 \right] u,
\end{equation}

we can simply use the known forms of the nucleon form factors to calculate these contributions. In the above, $F_{1,2}^j (k^2)$ are the nucleon form factors for light quark flavor $j$, $G_A^j (k^2)$ is the nucleon axial form factor and $u$ is a nucleon spinor. The squared momentum transfer is $k^2$. As will be explained in the next section, use of the full form factors is not necessary in the low momentum transfer limit which is relevant for nuclear recoil cross sections, and an effective nucleon current can be constructed using only the valence quarks.

To evaluate the heavy quark term, we use the heavy quark expansion as detailed in \cite{Franz:2000ee}. The heavy quark expansion allows us to reframe the heavy quark degrees of freedom in the nucleon in terms of the light quark and gluon degrees of freedom. This is useful because the latter are experimentally accessible, while the former are not. We note that there is an alternative method to the heavy quark expansion that has been used to extract heavy quark contributions to the proton axial vector currents \cite{Bass:2002mv}.

The heavy quark expansion of the vector current is zero to leading order, with nonzero contributions proportional to $1/m_Q^4$:

\begin{equation}
\langle N | \bar{Q}\gamma^\mu Q | N \rangle = 0 + \mathcal{O}\left( \frac{1}{m_Q^4}\right).
\end{equation}

We neglect these terms, and take the vector part of the current to be zero.

The heavy quark expansion of the axial vector contribution can be related to the expansion of the pseudoscalar current \cite{Franz:2000ee}:

\[
\langle N | \bar{Q} \gamma^\mu \gamma^5 Q | N \rangle = \frac{-ig_s^2}{96\pi^2 m_Q^2}\langle N |\left( \partial^\mu \mathrm{Tr}_c \left[ t^a G^a_{\alpha\beta}t^b\tilde{G}^{b\alpha\beta}\right]\right. \]
\begin{equation}\label{PSexp}
\left. + 4 \mathrm{Tr}_c \left[ \left[ D_\alpha , t^aG^{a\alpha\nu}\right] t^b\tilde{G}^{b\mu}_\nu \right] \right)| N \rangle + \mathcal{O}\left( \frac{1}{m_Q^4} \right).
\end{equation}

The gluon field strength tensor is $G_{\mu\nu}^a=\partial_\mu B^a_\nu - \partial_\nu B^a_\mu + g_s f^{abc}B^b_\mu B^c_\nu$ and the dual field strength tensor is $\tilde{G}_{\mu\nu}=(1/2)\epsilon^{\alpha\beta\mu\nu}G_{\mu\nu}$. The trace is over colors and $D_\alpha$ is the $SU(3)_c$ gauge covariant derivative.

The first term is simply proportional to the gluonic pseudoscalar nucleon matrix element $\langle N | G^a_{\alpha\beta}\tilde{G}^{a\alpha\beta} | N \rangle$ multiplied by the momentum transfer $k_\mu$, which is apparent after a Fourier transform and the exploitation of translation invariance. The first term becomes

\begin{equation}\label{HQ1}\frac{-ig_s^2}{96\pi^2 m_Q^2}\langle N | \partial^\mu \mathrm{Tr}_c \left[ t^aG^a_{\alpha\beta}t^b\tilde{G}^{b\alpha\beta}\right] | N \rangle = \frac{-i \alpha_s k^\mu}{48\pi m_Q^2} \langle N | G^a_{\alpha\beta}\tilde{G}^{a\alpha\beta} | N \rangle.
\end{equation}

Numerical values for the matrix element appear in the literature \cite{Cheng:2012qr} ($\sim 380$ MeV for the proton and $\sim -11$ MeV for the neutron). This contribution is suppressed by the momentum transfer and a factor of $1/m_Q^2$.

We now attack the second term in equation \ref{PSexp}. By using the QCD equation of motion $\left[ D_\alpha , G^{\alpha\nu}\right] = -i g_s^2  \sum_{q=u,d,s} \bar{q} \gamma^\nu q$, the second term can be written

\begin{equation}
\frac{ig_s^2}{24\pi^2 m_Q^2}\langle N | \mathrm{Tr}_c \left[ \left[ D_\alpha , t^aG^{a\alpha\nu}\right] t^b\tilde{G}^{b\mu}_\nu \right] | N \rangle =  \frac{i\alpha_s}{24\pi m_Q^2} \langle N | g_s \sum_{q=u,d,s} \bar{q} \gamma_\nu \tilde{G}^{\mu\nu} q | N \rangle .
\end{equation}

The right hand side of this equation can be expanded in terms of its Lorentz structure into terms proportional to the momentum transfer $k^\mu$ and the nucleon spin $S^\mu$:

\begin{equation}
\sum_{q=u,d,s}\langle N | g_s \bar{q} \gamma_\nu \tilde{G}^{\nu\mu} q | N \rangle = F_1 k^\mu + F_2 S^\mu.
\end{equation}

The term proportional to nucleon spin can be evaluated to get in the forward matrix element limit that corresponds to low momentum transfer \cite{Zhang:2012da,Ji:1997gs}

\begin{equation}
F_2 S^\mu = \sum_{q=u,d,s} 2 f_{N2,q}m_N^2S^\mu.
\end{equation}

The coefficients $f_{N2,q}$ are estimated to sum to 0.1 over the light flavors \cite{Ji:1997gs}.

Applying the Dirac equation to the quark fields in the momentum proportional term gives $\partial_\mu q = (m_q /4) \gamma_\mu q$ to leading order in fixed-point gauge (further details in the next Section), and subsequent use of the anticommutation relation for Dirac gamma matrices $\{\gamma_\mu , \gamma_\nu \} =2g_{\mu\nu}$ simplifies the expression to

\begin{equation}\label{HQ1}
F_1 =\sum_{q=u,d,s} \frac{m_q}{2}  \langle N | g_s \bar{q} \tilde{G}^{\nu\mu} q | N \rangle g_{\nu\mu} =0. 
\end{equation}

This quantity vanishes due to the antisymmetry of $\tilde{G}^{\nu\mu}$. The spin proportional term is not zero, and cannot be discarded out of hand when bottom quarks are taken into account. We find that the dominant part of the heavy quark contribution is 

\begin{equation}\label{HQdom}
\sum_Q \langle N | \bar{Q}\gamma^\mu \left( V^\prime_Q + A^\prime_Q \gamma^5 \right) Q | N \rangle \simeq \sum_Q A^\prime_Q \frac{4i m_N^2 \alpha_s \sum_{q=u,d,s} f_{N2,q}}{9\pi m_Q^2}S^\mu.
\end{equation}

Further discussion about how this term compares to the light quark contribution appears in the next Section.

Calculation of the gluon contribution is slightly more subtle. We use a 1-loop form of the $Z^\prime$-gluon-gluon interaction vertex $\Gamma_{Z^\prime gg}^{\mu\nu\rho} \left( p_1 ,p_2 \right)$ adapted from \cite{Duerr:2015wfa}. The vertex function depends on the gluon momenta $p_1$ and $p_2$, and has a cumbersome expression which we avoid reproducing in full generality. The crossed diagram where $p_1 \leftrightarrow p_2$ also needs to be taken into account, but as our calculations are symmetric in the gluon momenta, this simply results in an overall factor of 2. We note that this contribution is independent of the heavy quark vector coupling $V^\prime_Q$, depending only on the axial coupling $A^\prime_Q$. Consequently, if there is no heavy quark $U(1)^\prime$ axial charge, this contribution vanishes. A possible alternative to our treatment is to use an effective operator, as was done in \cite{Cox:2015afa}. 

To reduce the gluon contribution

\begin{equation}\label{gluoncont}
J_{Z^\prime NN}^{G,\mu}Z^\prime_\mu = \langle N | \Gamma^{\mu\nu\rho} \left( p_1 ,p_2 \right) Z^\prime_\mu \left( k \right) B_\nu \left( p_1 \right) B_\rho \left( p_2 \right) | N \rangle
\end{equation}

to a more useful form, we first Fourier transform the expectation value to position space, associating $\{k,p_1,p_2\} \sim \{x,y,z\}$. By working in fixed-point gauge, we can expand the gluon fields (inside the nucleon) near the origin in terms of their field strength tensors \cite{Cronstrom:1980hj}, yielding to leading order

\begin{equation}
B_\nu \left( y \right) B_\rho \left( z \right) \simeq \frac{y^\delta}{2} G_{\delta\nu}(0) \frac{z^\lambda}{2} G_{\lambda\rho}(0).
\end{equation}

At this point we have cast the color factors from the vertex function into the field strength tensors and suppressed the color structure indices, indicating this by the factorization $\Gamma_{Z^\prime gg}^{\mu\nu\rho} \left( p_1 ,p_2 \right)$ $= C \tilde{\Gamma}_{Z^\prime gg}^{\mu\nu\rho} \left( p_1 ,p_2 \right)$. We now have

\begin{equation}
J_{Z^\prime NN}^{G,\mu}Z^\prime_\mu \simeq \int d^4 x d^4 y d^4 z \langle N | \tilde{\Gamma}^{\mu\nu\rho} \left( y ,z \right) Z^\prime_\mu \left( x \right) \frac{y^\delta}{2} G_{\delta\nu}(0) \frac{z^\lambda}{2} G_{\lambda\rho}(0) | N \rangle e^{ik\cdot x}e^{ip_1\cdot y}e^{ip_2\cdot z}.
\end{equation}

Fourier transforming back to momentum space, the position coordinates become momentum derivatives and by exploiting symmetry properties and conservation of momentum ($p_1^\mu+p_2^\mu=k^\mu$), we can rearrange:

\begin{equation}
J_{Z^\prime NN}^{G,\mu}Z^\prime_\mu \simeq \frac{1}{4}\langle N | \tilde{\Gamma}^{\mu\nu\rho} \left( p_1 ,p_2 \right) \left( \frac{\partial}{\partial k^\delta} - \frac{\partial}{\partial k^\lambda} \right) Z^{\prime}_\mu \left( k \right) G^\delta_{\nu}(0) G^{\lambda}_{\rho}(0) | N \rangle .
\end{equation}

The vertex function $\tilde{\Gamma}$ and the derivatives can be pulled out of the expectation value. Contracting indices and integrating by parts allows the derivatives to be combined and simplifies the modified vertex function $\tilde{\Gamma}^{\mu\nu\rho} \left( p_1 ,p_2 \right)$ considerably, giving an overall factor

\begin{equation}
J_{Z^\prime NN}^{G,\mu}Z^\prime_\mu \simeq \sum_{Q} A_1^Q \frac{\partial}{\partial k^\delta} Z^{\prime}_{\mu} \left( k \right) \langle N | \tilde{G}^{\mu}_{\rho} G^{\delta\rho} | N \rangle .
\end{equation}

where the sum is over fermions contributing to the loop (the top quark for the top-philic case or the top and bottom quarks for the tritophilic case). An epsilon tensor was absorbed from the vertex function into the $\tilde{G}^{\mu}_{\rho}$. The extra factor of 2 comes from the contribution of the crossed diagram with gluon momenta exchanged. The matrix element $\langle N | \tilde{G}^{\mu}_{\rho} G^{\delta\rho} | N \rangle$ is the spin 2, twist 2 gluonic pseudotensor operator contribution to the nucleon matrix element. We note that assuming a vacuum spin structure for the gluons at this point would make the amplitude zero, in agreement with the Landau-Yang theorem \cite{Landau:1948kw,Yang:1950rg}. That this contribution is not zero is a consequence of the gluons being part of a hadronic bound state. 

The function $A_1^Q$ takes the form

\begin{equation}
A_1^Q = \frac{A^\prime_Q N_c}{4\pi^2}\left[ 3 + \Lambda (s, m_Q, m_Q) + 2 m_Q^2 C_0 (0, 0, s ;m_Q, m_Q, m_Q)\right],
\end{equation}

where the loop functions $\Lambda (s, m_Q, m_Q)$ and $C_0 (0, 0, s ;m_Q, m_Q, m_Q)$ are given explicitly in the reference \cite{Duerr:2015wfa}. Due to the structure of the triangle loop, only the axial charge contributes. For a zero $U(1)^\prime$ axial charge on the heavy quarks in the loop, the gluon contribution is thus zero. The variable $s=(p_1+p_2)^2$ takes the role of momentum transfer in our application (formally, crossing symmetry applies), and so we are interested in the low momentum transfer limit where $s=k^2 \rightarrow 0$. In the full limit $A_1^f \rightarrow 0$, but we would like to retain dependence on momentum transfer to the first order to be able to compare against other contributions to the elastic scattering matrix element.

Expanding the function $A_1^f$ around the point $k^2=0$ gives

\begin{equation}
A_1^f = \left(\frac{A^\prime_Q N_c}{4\pi^2}\right)\left(\frac{k^2}{12m_Q^2}\right) + \mathcal{O}\left( k^4 \right).
\end{equation}

The contribution from gluons is then for small momentum transfer

\begin{equation}\label{gluedom}
J_{Z^\prime NN}^{G,\mu}Z^\prime_\mu \simeq \frac{\partial}{\partial k^\delta} Z^\prime_\mu (k) \frac{A^\prime_Q}{16\pi^2m_Q^2}k^2  \langle N | \tilde{G}^{\mu}_{\rho} G^{\delta\rho} | N \rangle .
\end{equation}

This expression has unpleasant spin structure, but fortunately it is suppressed by the heavy quark mass squared and two powers of momentum transfer. It will clearly contribute subdominantly to a low momentum transfer scattering matrix element.

We note that, in contrast to the Higgs-gluon case \cite{shifman1978,cheng1988} which contributes significantly to the Higgs-nucleon interaction, the $Z^\prime$-gluon vertex does not include a large $Z^\prime$-top coupling to counteract the $1/m_Q^2$ suppression (this was also noted in \cite{Cox:2015afa}).

\section{Nuclear Recoil Cross Sections}

These results are most useful when considered in the context of nuclear recoil cross sections for the direct detection of dark matter. The cross sections for the scattering of fermionic (or scalar) vector-portal dark matter off of an atomic nucleus are cumbersome and appear elsewhere in the literature (i.e. \cite{Berlin:2014tja}), so we do not reproduce them here. 

A matrix element for dark matter-nucleon scattering mediated by a $Z^\prime$ has the generic form

\begin{equation}
\mathcal{M}\sim \mathcal{F}_{\mu (\nu )} (k^2) \left[ \mathcal{O}_{DM} \right] ^{(\nu ) (5)} \bar{u} \left[ J^\mu_{Z^\prime NN} \right] u
\end{equation}

where $\mathcal{F}_{\mu (\nu )} (k^2)$ is some form factor, the factor $\bar{u} \left[ J^\mu_{Z^\prime NN} \right] u$ describes the $Z^\prime$-nucleon interaction and the operator $\mathcal{O}_{DM}$, which may be scalar, vector, axial vector, pseudoscalar, or some combination, describes the dark matter interaction with the $Z^\prime$. In the nonrelativistic limit of interest for dark matter direct detection, whether or not this matrix element generates nuclear recoil cross sections that are dependent on nuclear spin or independent of it is due to the structure of the above operators. Additionally, some contributions may be suppressed by one or more powers of the momentum transfer, which is small in the nonrelativistic limit. For a detailed analysis of the matrix elements of dark matter nuclear recoils, see \cite{Kumar:2013iva}. 

First, we discuss the light quark contribution. The form factor description is not entirely necessary for the low momentum transfer limit, meaning we can write an effective nucleon current 

\begin{equation}\label{effnuccur}
J^\mu_{Z^\prime NN, q} = \bar{N} \gamma^\mu \left( V^\prime_N + A^\prime_N \gamma^5 \right) N.
\end{equation}

The nucleon effective charges are simply the summed contributions of the valence quark charges (and so will differ between the neutron and proton). This is the usual method for treating nucleon interactions in vector-portal models, and most cross section results reported in the literature are framed in these terms. We point out that this neglects the strange quark contribution, which we consider acceptable. Depending on the form of the dark matter operator, the light quark contribution can contain in general both spin dependent and spin independent terms, as well as terms suppressed by the momentum transfer. Making further statements requires choosing a specific model.

This term is proportional only to $g_{mix}$, with no suppression from heavy quark masses like the other terms, and is the dominant contribution to the matrix element unless the $U(1)^\prime$ gauge coupling $g^\prime$ is large enough when compared to the mixed coupling $g_{mix}$ that it can overcome the suppression. However, $g_{mix}$ also depends on $g^\prime$, introducing complications. In the following, we consider the conditions that might lead to this reversal.

The gluon contribution \ref{gluedom} is proportional not only to the gluonic pseudotensor matrix element $ \langle N |O^{2\mu\nu}_{5g} | N \rangle$, which makes it spin dependent, but is also suppressed by (two) powers of the momentum transfer and a factor of $1/m_Q^2$. This pseudotensor matrix element is rarely discussed in the literature \cite{Solon:2016yai}, but it is proportional to the nucleon mass and spin. No other contribution to the matrix element is so suppressed, so we feel comfortable in neglecting the gluon term. Interference terms would also contribute negligibly in the nonrelativistic limit \cite{Kumar:2013iva}.

The dominant term \ref{HQdom} from the heavy quark contribution is the largest of the non-light quark contributions. It is suppressed only by a factor of $1/m_Q^2$, and generates a spin-dependent contribution to the nuclear recoil matrix element. It might seem safe to ignore this term as well, but as it is the largest of the subdominant terms, we should examine what $U(1)^\prime$ gauge coupling is required for this term to become comparable to the light quark contributions. If there were zero mixing, the light quark contributions would be zero and the heavy quark contribution \ref{HQdom} would dominate. With mixing, things are different. A basic analysis reveals that \ref{HQdom} becomes relevant roughly when

\begin{equation}\label{gaugecomp}
\frac{g^\prime}{600\pi m_Q^2}\simeq g_{mix}.
\end{equation}

If only the top quark is included, this requires approximately $g^\prime \simeq 1.8\times 10^7 \ g_{mix}$ while if the bottom quark is involved, the requirement is $g^\prime \simeq 1\times 10^4 \ g_{mix}$. These are unrealistic requirements for a $U(1)^\prime$ gauge parameter unless $g_{mix}$ is very small. However, as we discussed in Section 2, quantum effects introduce mixing proportional to $g^\prime$.

In fact, one can use the approximation for loop-induced mixing in equation \ref{loopmix} and the expressions for $g_{mix}$ in \cite{Chankowski:2006jk} to obtain a somewhat complicated equation for $g^\prime$ that can be solved to give the regions where equation \ref{gaugecomp} is satisfied. The exact relationship depends on the symmetry breaking pattern and the new scales involved, as well as contributions from both mass and kinetic mixing. We tested a basic case with a generic symmetry breaking pattern ($SU(2)_L\times U(1)_Y \times U(1)^\prime \rightarrow SU(2)_L\times U(1)_Y \rightarrow U(1)_{em}$), loop induced kinetic mixing, no mass mixing, and a scale of $\Lambda_{UV}\sim\mathcal{O} (\mathrm{TeV} )$. In this unrealistic and oversimplified scenario we found that the equation had no real solutions; that is, there was no value of $g^\prime$ such that the heavy quark contribution was comparable to the light quark contribution. 

Of course, an individual model would need to be examined in detail before such a claim could be made. Because the heavy quark contribution is spin-dependent, and the spin-dependent exclusion bounds are significantly weaker than the spin-independent bounds \cite{Akerib:2016lao,Amole:2015pla}, the charges would have to be carefully tuned to set the spin-independent part of the cross section to zero for the heavy quark term to have any relevance. We note as well that values of 0.01 are not unreasonable for a tree level $g_{mix}$ in realistic models \cite{Wang:2015sxe}. For more complicated symmetry breaking scenarios or situations with exceptionally small gauge couplings, we urge a cautious treatment of the effects of mixing. In general, we argue that the mixing induced contributions from light quarks are dominant and that the effective nucleon current \ref{effnuccur} can be applied.

Finally, we note that there could be contributions to the matrix element from SM $Z$ boson exchange. In many scenarios it is possible to perform the gauge field rotations in such a way that the $Z$ boson does not acquire a mixed current coupled to $U(1)^\prime$ charge \cite{Chankowski:2006jk}, but this is not the case in general. If the $Z$ obtains a coupling to the dark matter sector, $Z$ exchange contributions to the matrix element need to be taken into account. The nucleon coupling is simply a modification of the current in equation \ref{effnuccur}, only with SM neutral currents, while the dark matter current will be proportional to $g_{mix}$. Whether or not these terms are relevant depends on the mass hierarchy of the $Z$ and $Z^\prime$, as well as how the gauge couplings $g^\prime$ and $g_Y$ compare. Again, we urge careful consideration of what basis the interacting gauge fields are in.

\section{Summary and Conclusions}

This article has provided a vertex function for low momentum transfer interactions between the nucleon and a top-philic or tritophilic $Z^\prime$ boson. We have taken into account contributions from light quarks induced by mixing, from heavy quarks, and from the gluonic content of the nucleon. We have found that for most realistic scenarios, the light quark contribution dominates. The gluonic contribution is always subdominant, but the heavy quark term could potentially contribute. 

Implications for the direct detection of top-philic or tritophilic vector-portal dark matter were considered. Exact results depend on the dark matter model, but with explicit tree level mixing included the light quark term will dominate unless the mixed coupling $g_{mix}$ is 4-8 orders of magnitude smaller than the $U(1)^\prime$ gauge coupling $g^\prime$. Due to quantum corrections to the mixed coupling, there may be no value where the heavy quark contribution proportional to $g^\prime$ becomes relevant.

We have also discussed the conditions in which mixing effects can be ignored in top-philic or tritophilic models with a $Z^\prime$. While it is difficult to draw model-specific conclusions, for low momentum transfer processes mixing effects cannot in general be ignored. Careful analysis is required before simplifying assumptions to ignore mixing effects can be made in non-universal models.

\subsubsection*{Acknowledgements}

This work was supported in part by the Natural Sciences and Engineering Research Council of Canada (NSERC) Discovery Grant Program.

\bibliographystyle{h-physrev_mod}
\bibliography{TopphilZ}

\begin{thebibliography}{10}

\bibitem{Pospelov:2007mp}
M.~Pospelov, A.~Ritz, and M.~B. Voloshin,
\newblock Phys. Lett. {\bf B662}, 53 (2008), 0711.4866.

\bibitem{Alves:2013tqa}
A.~Alves, S.~Profumo, and F.~S. Queiroz,
\newblock JHEP {\bf 04}, 063 (2014), 1312.5281.

\bibitem{Jacques:2016dqz}
T.~Jacques {\em et~al.},
\newblock (2016), 1605.06513.

\bibitem{Dudas:2012pb}
E.~Dudas, Y.~Mambrini, S.~Pokorski, and A.~Romagnoni,
\newblock JHEP {\bf 10}, 123 (2012), 1205.1520.

\bibitem{Chun:2010ve}
E.~J. Chun, J.-C. Park, and S.~Scopel,
\newblock JHEP {\bf 02}, 100 (2011), 1011.3300.

\bibitem{Cline:2014dwa}
J.~M. Cline, G.~Dupuis, Z.~Liu, and W.~Xue,
\newblock JHEP {\bf 08}, 131 (2014), 1405.7691.

\bibitem{Alves:2015mua}
A.~Alves, A.~Berlin, S.~Profumo, and F.~S. Queiroz,
\newblock JHEP {\bf 10}, 076 (2015), 1506.06767.

\bibitem{Cox:2015afa}
P.~Cox, A.~D. Medina, T.~S. Ray, and A.~Spray,
\newblock JHEP {\bf 06}, 110 (2016), 1512.00471.

\bibitem{Arina:2016cqj}
C.~Arina {\em et~al.},
\newblock (2016), 1605.09242.

\bibitem{PhysRevD.91.035025}
D.~Hooper,
\newblock Phys. Rev. {\bf D91}, 035025 (2015).

\bibitem{Beck:2015cga}
L.~Beck {\em et~al.},
\newblock Phys. Lett. {\bf B746}, 48 (2015), 1501.07580.

\bibitem{Hill:1994hp}
C.~T. Hill,
\newblock Phys. Lett. {\bf B345}, 483 (1995), hep-ph/9411426.

\bibitem{Lane:1995gw}
K.~D. Lane and E.~Eichten,
\newblock Phys. Lett. {\bf B352}, 382 (1995), hep-ph/9503433.

\bibitem{Lane:1996ua}
K.~D. Lane,
\newblock Phys. Rev. {\bf D54}, 2204 (1996), hep-ph/9602221.

\bibitem{Lane:1998qi}
K.~D. Lane,
\newblock Phys. Lett. {\bf B433}, 96 (1998), hep-ph/9805254.

\bibitem{Popovic:1998vb}
M.~B. Popovic and E.~H. Simmons,
\newblock Phys. Rev. {\bf D58}, 095007 (1998), hep-ph/9806287.

\bibitem{He:1999vp}
H.-J. He, T.~M.~P. Tait, and C.~P. Yuan,
\newblock Phys. Rev. {\bf D62}, 011702 (2000), hep-ph/9911266.

\bibitem{Holdom:1994ss}
B.~Holdom,
\newblock Phys. Lett. {\bf B339}, 114 (1994), hep-ph/9407311.

\bibitem{Frampton:1996bs}
P.~H. Frampton, M.~B. Wise, and B.~D. Wright,
\newblock Phys. Rev. {\bf D54}, 5820 (1996), hep-ph/9604260.

\bibitem{Akerib:2015rjg}
LUX Collaboration, D.~S. Akerib {\em et~al.},
\newblock Phys. Rev. Lett. {\bf 116}, 161301 (2016), 1512.03506.

\bibitem{Tan:2016zwf}
PandaX-II Collaboration, A.~Tan {\em et~al.},
\newblock Phys. Rev. Lett. {\bf 117}, 121303 (2016), 1607.07400.

\bibitem{shifman1978}
M.~Shifman, A.~Vainshtein, and V.~Zakharov,
\newblock Phys. Lett. {\bf B78}, 443  (1978).

\bibitem{cheng1988}
T.~P. Cheng,
\newblock Phys. Rev. {\bf D38}, 2869 (1988).

\bibitem{Novikov:1977dq}
V.~A. Novikov {\em et~al.},
\newblock Phys. Rept. {\bf 41}, 1 (1978).

\bibitem{delAguila:1988jz}
F.~del Aguila, G.~D. Coughlan, and M.~Quiros,
\newblock Nucl. Phys. {\bf B307}, 633 (1988),
\newblock [Erratum: Nucl. Phys.B312,751(1989)].

\bibitem{Chankowski:2006jk}
P.~H. Chankowski, S.~Pokorski, and J.~Wagner,
\newblock Eur. Phys. J. {\bf C47}, 187 (2006), hep-ph/0601097.

\bibitem{Foot:1991kb}
R.~Foot and X.-G. He,
\newblock Phys. Lett. {\bf B267}, 509 (1991).

\bibitem{Holdom:1985ag}
B.~Holdom,
\newblock Phys. Lett. {\bf B166}, 196 (1986).

\bibitem{Berlin:2014tja}
A.~Berlin, D.~Hooper, and S.~D. McDermott,
\newblock Phys. Rev. {\bf D89}, 115022 (2014), 1404.0022.

\bibitem{Greiner:2014qna}
N.~Greiner, K.~Kong, J.-C. Park, S.~C. Park, and J.-C. Winter,
\newblock JHEP {\bf 04}, 029 (2015), 1410.6099.

\bibitem{Kim:2016plm}
J.~H. Kim, K.~Kong, S.~J. Lee, and G.~Mohlabeng,
\newblock Phys. Rev. {\bf D94}, 035023 (2016), 1604.07421.

\bibitem{Gresham:2011dg}
M.~I. Gresham, I.-W. Kim, and K.~M. Zurek,
\newblock Phys. Rev. {\bf D84}, 034025 (2011), 1102.0018.

\bibitem{Gupta:2010wt}
S.~K. Gupta and G.~Valencia,
\newblock Phys. Rev. {\bf D82}, 035017 (2010), 1005.4578.

\bibitem{Franz:2000ee}
M.~Franz, M.~V. Polyakov, and K.~Goeke,
\newblock Phys. Rev. {\bf D62}, 074024 (2000), hep-ph/0002240.

\bibitem{Bass:2002mv}
S.~D. Bass, R.~J. Crewther, F.~M. Steffens, and A.~W. Thomas,
\newblock Phys. Rev. {\bf D66}, 031901 (2002), hep-ph/0207071.

\bibitem{Cheng:2012qr}
H.-Y. Cheng and C.-W. Chiang,
\newblock JHEP {\bf 07}, 009 (2012), 1202.1292.

\bibitem{Zhang:2012da}
Y.~Zhang,
\newblock Phys. Lett. {\bf B720}, 137 (2013), 1212.2730.

\bibitem{Ji:1997gs}
X.-D. Ji and W.~Melnitchouk,
\newblock Phys. Rev. {\bf D56}, R1 (1997), hep-ph/9703363.

\bibitem{Duerr:2015wfa}
M.~Duerr, P.~Fileviez~Perez, and J.~Smirnov,
\newblock Phys. Rev. {\bf D92}, 083521 (2015), 1506.05107.

\bibitem{Cronstrom:1980hj}
C.~Cronstrom,
\newblock Phys. Lett. {\bf B90}, 267 (1980).

\bibitem{Landau:1948kw}
L.~D. Landau,
\newblock Dokl. Akad. Nauk Ser. Fiz. {\bf 60}, 207 (1948).

\bibitem{Yang:1950rg}
C.-N. Yang,
\newblock Phys. Rev. {\bf 77}, 242 (1950).

\bibitem{Kumar:2013iva}
J.~Kumar and D.~Marfatia,
\newblock Phys. Rev. {\bf D88}, 014035 (2013), 1305.1611.

\bibitem{Solon:2016yai}
M.~P. Solon,
\newblock {\em {Heavy WIMP Effective Theory}},
\newblock PhD thesis, UC, Berkeley, Cham, 2014.

\bibitem{Akerib:2016lao}
LUX Collaboration, D.~S. Akerib {\em et~al.},
\newblock Phys. Rev. Lett. {\bf 116}, 161302 (2016), 1602.03489.

\bibitem{Amole:2015pla}
PICO Collaboration, C.~Amole {\em et~al.},
\newblock Phys. Rev. {\bf D93}, 052014 (2016), 1510.07754.

\bibitem{Wang:2015sxe}
Z.-W. Wang, F.~S. Sage, T.~G. Steele, and R.~B. Mann,
\newblock (2015), 1511.02531.

\end{thebibliography}

\end{document}